# Chiral lone-pair helices with handedness coupling to electric-strain fields

*Caren R. Zeiger#, Ruben S. Dragland, Robin Sjökvist, Richard Beanland, Dennis Meier, Tor Grande, Mark S. Senn and Ola G. Grendal\**

C. R. Zieger, R. S. Dragland, D. Meier, T. Grande, O. G. Grendal
Department of Materials Science and Engineering, NTNU Norwegian University of Science and Technology, Sem Sælands vei 12, 7034 Trondheim, Norway
E-mail: ola.g.grendal@ntnu.no

R. Sjökvist, R. Beanland
Department of Physics, University of Warwick, Gibbert Hill Road, Coventry CV4 7AL, United Kingdom

M. S. Senn
Department of Chemistry, University of Warwick, Gibbert Hill Road, Coventry CV4 7AL, United Kingdom

# Current affiliation: SINTEF Industry, Richard Birkelands vei 3, 7034 Trondheim, Norway

Funding: O.G.G acknowledges financial support from NTNU (Norwegian University of Science and Technology). C.R.Z. and T.G. acknowledge funding from the Norwegian Research Council through the project High-Temperature Lead-Free Ferroelectrics based on Tungsten Bronzes (301954). R.S.D. and D.M. acknowledge funding from the European Research Council (ERC) under the European Union's Horizon 2020 Research and Innovation Program (Grant Agreement No. 863691). R.S acknowledges funding from EPSRC grant EP/V053701/1. M.S.S acknowledges the Royal Society for a fellowship (URF\R\231012).

**Abstract:** Ferrochiral materials with an achiral-to-chiral phase transition and switchable chirality have unique application opportunities, enabling control of the angular momentum of circularly polarized lattice vibrations (chiral phonons) and chirality-related electronic phenomena. Materials that fall into this class are, however, extremely rare, and often accompanied by other types of ferroic order that interfere with the ferrochiral responses. In this work, we demonstrate ferrochirality in two tetragonal tungsten bronzes, $K_4Bi_2Nb_{10}O_{30}$ and $Rb_4Bi_2Nb_{10}O_{30}$. Using high-resolution X-ray powder diffraction combined with transmission electron microscopy, we solve the incommensurately modulated and chiral structures. Temperature dependent X-ray powder diffraction reveals that both materials undergo an achiral-to-chiral phase transition from $P4/mbm$ to $P42_12(00\gamma)q00$. The chirality originates from a cooperative helical displacement of $Bi^{3+}$ atoms perpendicular to the $c$ direction and represents the primary order parameter. As a secondary effect of the ferrochiral order, a spatially varying piezoelectric response is observed, consistent with the polycrystalline nature of the investigated materials. Through invariant analysis, an external electric-strain-field coupling with the piezoelectricity is proposed as a conjugate field for switching chirality, establishing tetragonal tungsten bronzes as a versatile playground for the emergent field of ferrochirality.

# 1. Introduction

Chirality is a crucial and fundamental concept in natural sciences, having significant implications in fields ranging from cosmology to particle physics, as well as chemistry, medicine and biology [1, 2]. In solid-state physics, chirality is closely linked to optical activity [3] and is gaining increasing interest due to topics like magnetic skyrmions,[4] topological insulators,[5] magneto-chiral dichroism,[6] chiral quantum optics, [7] and negative refractive index for "perfect" lenses.[8] Following from this, there is a growing interest in direct external-field control of chirality, or so-called *ferrochiral* materials, defined as materials with an achiral-to-chiral phase transition and switchable chirality. The ferrochirality could, for example, enable direct control and switching of optical activity in contrast to the indirect switching, achieved through a coupling to, *e.g.*, ferroelectricity [9] or ferroelasticity.[10] Still, chirality is relatively unexplored in the field of solid-state physics and inorganic chemistry,[11] partly due to the scarcity of potential ferrochiral material candidates, with only a few well-established examples.[12-15]

In general, structural chirality in periodic inorganic solids can manifest in one of two ways. It either occurs through assembly of *achiral* building blocks in one of the 22 *enantiomorphic* Sohncke space groups, or assembly of *chiral* building blocks in one of the 43 *non-enantiomorhphic* Sohncke space groups.[11] Despite chiral building blocks being more common in organic than in inorganic solids, there exist examples of chiral inorganic solids such as $Pb_5Ge_3O_{11}$ [16] and $Ba(TiO)Cu_4(PO_4)_4$,[17], which are described by the non-enantiomorphic space groups $P3$ and $P4_22$, respectively.

Recently, inducing helicity (*i.e.*, a chiral object/building block) as a way of inducing chirality in periodic inorganic materials has been performed in the one-dimensional van der Waals crystal, GaSeI.[18] Helicity was induced through face-sharing tetrahedra, forming a Boerdijk-Coxeter helix; however, the resulting structure was *antichiral* (*i.e.*, contains compensated opposite chirality in space group $Cc$), not chiral. A similar approach, although using atoms instead of tetrahedra as "building blocks", was utilized to successfully induce chirality in the Cu-doped quadruple perovskite $BiMn_7O_{12}$.[19] It was shown that $Bi^{3+}$ takes on an unusual atomic arrangement of helical "chains", reminiscent of helical magnetic spin arrangements, resulting in an incommensurate modulated structure described by super space group $R3(00\gamma)t$ ($R3$ being one of the 43 non-enantiomorphic space groups).[19] With this work, we pursue a similar avenue for inducing chirality as for $BiMn_7O_{12}$, that is, by introducing helicity in another perovskite related structure, namely that of tetragonal tungsten bronzes.

The tetragonal tungsten bronze (TTB) crystal structure is comparable to that of the known conventional perovskite structure with corner-sharing octahedra, but with a more complex general formula ($A1_2A2_4C_2B1_2B2_8O_{30}$) and aristotype space group $P4/mbm$ (no. 127).[20, 21] The structure is flexible both with respect to accommodating various cations across the A1-, A2- and C-sites and towards structural distortions.[22] These structural distortions are sometimes found to be incommensurate modulations,[22, 23] having a periodicity with an irrational relation to the underlying average lattice. This incommensurability breaks additional symmetries and, hence, provides extra flexibility when searching for novel structures and emergent properties.

Recently, an uncommon incommensurate modulation for TTBs, with a modulation vector of the form $k = [0, 0, \gamma]$, was reported for $K_4Bi_2Nb_{10}O_{30}$ (KBN) and $Rb_4Bi_2Nb_{10}O_{30}$ (RBN).[24] However, only their average structure was solved and found to be described by the aristotype $P4/mbm$ space group. Their incommensurate modulated structures are yet to be solved. Furthermore, it was shown that neither KBN nor RBN exhibit spontaneous electric order,[24] and their average tetragonal structure indicates that there is no improper ferroelastic coupling,[25, 26] classifying them as non-ferroelectric, non-ferroelastic systems.

In this work, we solve the unusual incommensurate modulated structure of KBN and RBN using a combination of high-resolution X-ray powder diffraction and scanning transmission electron microscopy (STEM). The observed structures are chiral, originating from a cooperative helical displacement of $Bi^{3+}$ atoms perpendicular to the $c$ direction. The helical displacements are described by a wave vector that is parallel to $c$. Both KBN and RBN show an achiral-to-chiral phase transition that is studied as a function of temperature. Symmetry analysis reveals that the mode related to chirality is the primary symmetry-breaking order parameter of the phase transition. Furthermore, "improper piezoelectricity" arises due to a secondary order parameter, which is experimentally verified using piezoresponse force microscopy. Invariant analysis suggests the possibility of "chirality-switching" in KBN and RBN, representing an intriguing pathway towards reversible chirality control in inorganic solids.

## 2. Results and discussion

### 2.1. Chiral incommensurate structure solution of KBN and RBN

The main reflections observed in the X-ray diffraction data for both KBN and RBN at room temperature, can be indexed by the average parent structure in $P4/mbm$ (KBN: $a$ = 12.64153(2) Å, $c$ = 3.92802(1) Å, RBN: $a$ = 12.72244(2) Å, $c$ = 3.95744(1) Å). In addition to these main reflections, clear 1st order satellite reflections are observed in the X-ray diffraction data (see orange tick marks in **Figure 1a**), which can be indexed with a modulation vector of the form [0, 0, γ] for both KBN ($k_{KBN}$ = [0, 0, 0.1379(2)]) and RBN ($k_{RBN}$ = [0, 0, 0.1405(1)]). No higher-order satellite reflections are resolved. Superspace group $P4_{2}12(00\gamma)q00$ (no. 90.1.19.2) with cell basis {(1,0,0), (0,1,0), (0,0,1)} and origin shift (1/2, 0, 0) compared to the $P4/mbm$ parent structure was found as the correct structural description (see Methods for further details on the structure solution). This finding was confirmed by Rietveld refinement of the high-resolution X-ray diffraction data, as shown in Figure 1a) for KBN and Figure S1 for RBN.

Superspace group Rietveld refinement revealed that the most striking feature of this structure is a helix, formed along the $c$ direction of the average tetragonal cell by the $Bi^{3+}$ atoms on the A1-site. The $Bi^{3+}$ atoms are displaced in-plane perpendicular to the $c$ direction, as shown for an approximate commensurate 1×1×7 supercell description in Figure 1b) and c). The displacement of $Bi^{3+}$ away from the centre position is 0.39 Å and 0.44 Å for KBN and RBN, respectively. More detailed crystallographic data of the room temperature structures of KBN and RBN are given in Table S1 and Table S2, respectively. $P4_{2}12$ is one of the 43 non-enantiomorphic Sohncke space groups, and the structure of KBN and RBN is chiral, driven by the helical arrangement of the $Bi^{3+}$ atoms. It is important to note that, for chiral structures described by non-enantiomorphic space groups, the "left-handed" and "right-handed" versions are described by the same space group, differing only in the "handedness" of the chiral object that is present, as shown in Figure 1c).

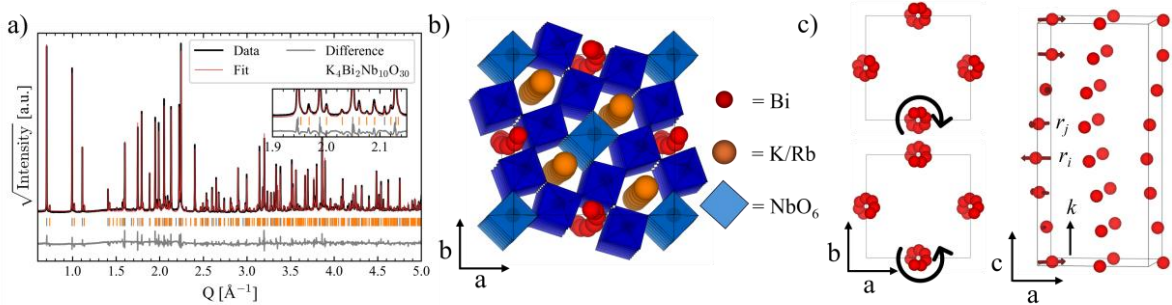

**Figure 1.** a) Room temperature high resolution X-ray powder diffraction data of KBN with Rietveld refinement using space group $P4_{2}12(00\gamma)q00$ and $k$ = [0, 0, 0.1379(2)]. Gray tick marks indicate main reflections (*i.e.*, [*hkl*0]), while the orange tick marks indicate satellite reflections (*i.e.*, [*hklm*], where $m$ = 1 for 1$^{st}$ order satellites). b) Commensurate supercell approximation of the incommensurate structure of KBN/RBN showing the helical displacement of $Bi^{3+}$ atoms perpendicular to the $c$ direction. c) Schematic highlighting the helical displacement of $Bi^{3+}$ atoms, including a top-down-view of the left and right-handed helix giving the two versions of the chiral structure. In c) the displacement has been exaggerated for clarity. Crystal structures made using *VESTA*.[27]

The incommensurate nature of KBN and RBN, and the unusual helical displacements of $Bi^{3+}$ atoms on the A1-site, can also be seen in real space. Atomic resolution STEM, shown in **Figure 2** and Figure S2 for KBN and RBN, respectively, clearly show the helical displacement. The images are taken along the [120] zone axis (the [120] projected average structure showing only $Bi^{3+}$ and $Nb^{5+}$ is overlaid in the inset) and show pronounced modulations for the $Bi^{3+}$ atoms perpendicular to the *c* direction (highlighted with red markers) with a period of ≈ 7 "atom rows". This modulation correlates well with the refined modulation vectors and the helical displacement determined from X-ray powder diffraction. Additional atomic displacement vector/field maps are presented in Figure S3 for KBN and RBN, which also highlight the helical displacement. Furthermore, selected area electron diffraction (SAED) patterns show clear satellite reflections (see Figure S4) that can be indexed with the incommensurate modulation vectors $k_{KBN}$ = [0, 0, 0.133(1)] and $k_{RBN}$ = [0, 0, 0.141(1)], in agreement with the values obtained from X-ray powder diffraction.

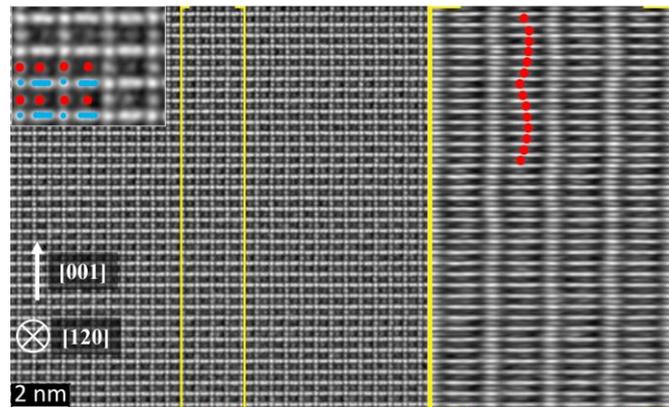

**Figure 2.** Atomic resolution STEM of KBN along the [120] zone axis. The inset shows the [120] projected average structure, highlighting only $Bi^{3+}$ (red) and $Nb^{5+}$ (blue), and the right panel is stretched to highlight the modulated displacement of $Bi^{3+}$ atoms.

We propose this unusual atomic arrangement to be an effect of a cooperative ordering of the stereochemical active lone-pair of $Bi^{3+}$. Off-centering of $Bi^{3+}$ (and other lone-pair containing cations) is frequently observed for perovskites, *e.g.*, in $PbTiO_3$ [28] and $BiFeO_3$,[29] and in the TTBs $Pb_5Nb_{10}O_{30}$ [30] and $Pb_2Bi_2Nb_{10}O_{30}$.[31] As mentioned, an incommensurate helical ordering of $Bi^{3+}$ atoms has also been observed in the Cu-doped quadruple perovskite $BiMn_7O_{12}$.[19] A common feature of all these materials is, however, that they are ferroelectric (or relaxor ferroelectric) as a direct consequence of this off-centering of the lone-pair cation, which separates them from the non-ferroelectric, non-ferroelastic KBN and RBN systems addressed in our work.[24]

For KBN, an additional set of incommensurate satellite reflections was observed in electron diffraction, see Figure S5, consistent with previous SAED reports.[24] Dark field TEM (see Figure S5) revealed that this second set of satellite reflections is associated with a complex domain structure with a length scale of tens of nanometers, comparable to other TTB systems.[32-35] However, these satellite reflections were not observed in the high-resolution X-ray diffraction data and are not further considered in the following.

**2.2. Nature of the achiral-to-chiral phase transition**

Having solved the room temperature incommensurate structure of KBN and RBN, we now focus on the temperature dependent structural changes and phase transition into the chiral state. At high temperatures (above 598 K and 723 K for KBN and RBN, respectively), the 1st order satellite reflections disappear, and the diffraction data is described by the aristotype *P4/mbm* space group (see Figure S6). Still, Rietveld refinement reveals a significant degree of positional disorder of the $Bi^{3+}$ atom on the A1-site, *i.e.*, similar in-plane off-centering as seen at room temperature for the $P4_12(00\gamma)q00$ structure (see Methods). It seems clear that the nature of this phase transition is that of an order-disorder phase transition. The $Bi^{3+}$ atoms are randomly off-centered above the phase transition temperature, resulting on average in a *P4/mbm* structure, and upon cooling order into a helical arrangement. No other phase transitions are observed down to 4K, making the chiral structure stable in a wide temperature range, including room temperature and well above. Given the chiral and incommensurate nature of the $P4_12(00\gamma)q00$ structure, this phase transition can be classified as commensurate-to-incommensurate and as achiral-to-chiral.

In a symmetry mode description,[36] the primary symmetry breaking order parameter ($Q_\Lambda$) of the phase transition from *P4/mbm* to $P4_12(00\gamma)q00$ is the chirality and transforms as the irreducible representation $\boldsymbol{\Lambda_5}$ with order parameter direction (*a*, 0, 0, 0). With this transition follows a single secondary order parameter ($Q_{piezo}$), transforming as the irreducible representation $\boldsymbol{\Gamma_1^-}$ with order parameter direction (*a*). High-resolution X-ray diffraction data was collected as a function of temperature from 4 to 723 K and 823 K for KBN and RBN, respectively, and Rietveld refinement with a symmetry mode approach was used to extract temperature dependent structural changes. The temperature evolution of the displacive distortion amplitudes associated with $\boldsymbol{\Lambda_5}$ and $\boldsymbol{\Gamma_1^-}$ is presented in Figure S7, and especially $\boldsymbol{\Lambda_5}$ shows a rapid increase below the phase transition before plateauing.

An alternative/derived primary order parameter of a helical structure can be defined in a similar way to what is done for helical magnetic spin orders, *i.e.*, using the mixed product $\boldsymbol{\sigma} =$ $\mathbf{k} \cdot [\mathbf{r_i} \times \mathbf{r_j}]$, where $k$ is the modulation vector and $r_i$ and $r_j$ are atomic positions in adjacent unit cells along the incommensurate propagation vector,[19] see Figure 1c). From the temperature dependent Rietveld refinements, we calculate $\sigma$ from the $Bi^{3+}$ positions, and as can be seen in **Figure 3**, $\sigma$ shows the same general trend as the $\Lambda_5$ mode, with a rapid increase below the phase transition before plateauing. X-ray powder diffraction data does not allow for determining the handedness of the structure (*i.e.*, both left- and right-handed helices, producing negative and positive values for $\sigma$, fit the data equally well). Furthermore, it is likely that our samples are *racemic*, showing a mix of both left- and right-handed versions of the helix. Thus, in this work, $\sigma$ should only be considered a measure of the "magnitude of chirality".

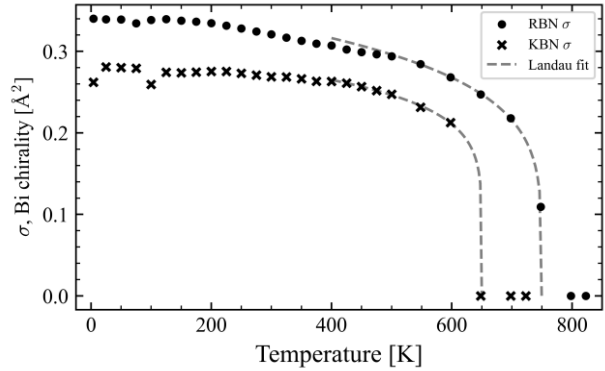

**Figure 3:** Temperature dependent $\sigma$ (see main text for definition) for KBN and RBN. Gray dashed line shows the critical exponent fit (see main text for details).

The temperature behaviour of $\Lambda_5$ and $\sigma$ is indicative of a second-order/continuous phase transition and, hence, expected to be proportional to the reduced temperature to the power of the critical exponent $\beta$, *e.g.*, $\boldsymbol{\sigma} \propto \left(\frac{|T-T_c|}{T_c}\right)^{\boldsymbol{\beta}}$ in the vicinity of the phase transition. We chose to fit $\sigma$ (based only on $Bi^{3+}$) instead of $\Lambda_5$ (based on all atoms in the structure), since our X-ray diffraction data is more sensitive to the $Bi^{3+}$ positions than oxygen positions, giving a clearer trend. Gray dashed lines in Figure 3 show the fits to the 5 first data points below the phase transition temperature (defined by visual inspection of the disappearance of the satellite reflections in the X-ray diffraction data). The obtained critical temperatures and critical exponents are 650 K and 0.14, and 749 K and 0.20, for KBN and RBN, respectively. The values for the critical exponent deviate from the theoretical value of 0.5 predicted by Landau theory, falling instead in-between the theoretical values for a 2D (0.125) and 3D (0.313) Ising model, comparable to values reported for antiferromagnetic phase transitions in $CuCl_2 \cdot 2H_2O$ and $CoCl_2 \cdot 6H_2O$ (0.15 – 0.29).[37] A possible explanation for the discrepancy is the lack of data points in the vicinity of the phase transitions, making the transition appear more first-order/discontinuous-like and preventing proper analysis by Landau theory.

The temperature evolution of the *a* and *c* lattice parameters is presented in **Figure 4** a). A close-to-linear decrease with decreasing temperature is observed, except a small plateau in the *a* lattice parameters around the phase transition temperature (marked with grey dashed lines), especially prominent for KBN. This behaviour agrees with previous work.[24] Neither the out-of-plane *c* or the in-plane *a*/*b* lattice parameters show the characteristic temperature dependence associated with out-of-plane [38, 39] or in-plane polarization [30] in TTBs. The modulation vector (Figure 4 b)) shows an initial decrease below the phase transition temperatures before plateauing, which this time is most prominent for RBN.

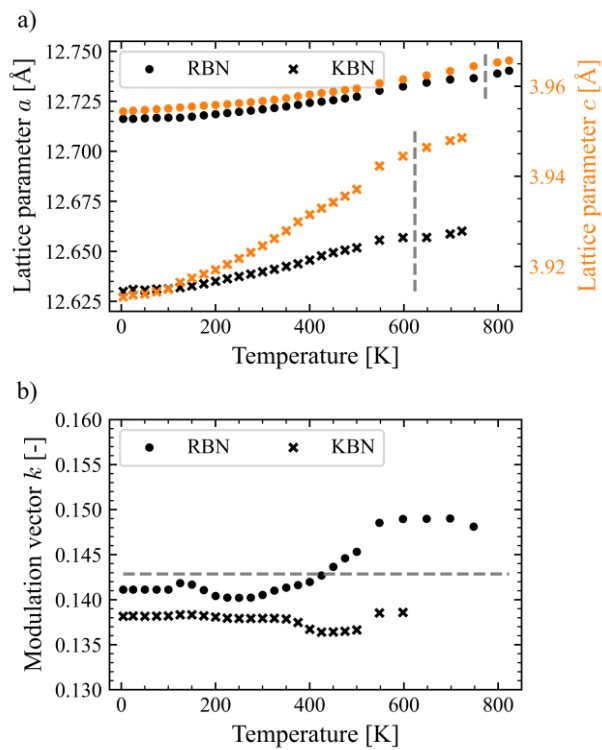

**Figure 4.** a) Temperature dependent *a* and *c* lattice parameters for KBN and RBN obtained by Rietveld refinement (grey dashed lines indicates the approximate phase transition temperatures). b) Temperature dependent modulation vector, *i.e.*, $\gamma$ in $k = [0, 0, \gamma]$, for KBN and RBN obtained by Rietveld refinement (grey dashed line indicates the commensurate value of 1/7).

In summary, an achiral-to-chiral second-order phase transition is observed for KBN and RBN at ~ 650 K and ~ 750 K, respectively, driven by $\Lambda_5$ or $\sigma$, which represent the primary symmetry breaking order parameter. Furthermore, considering the positional disorder observed in the high-temperature phase, the transition shows clear evidence of an order-disorder type transition.

## 2.3. Improper piezoelectricity in chiral KBN and RBN

The secondary order parameter ($Q_{piezo}$), which accompanies the achiral-to-chiral phase transition for KBN and RBN, transforms as $\Gamma_1^-$ and induces piezoelectricity. As $Q_{piezo}$ is not the primary order parameter, KBN and RBN can be classified as "improper piezoelectrics", by analogy with improper ferroelectrics.[40] We investigated this effect experimentally using piezoresponse force microscopy (PFM). Out-of-plane and in-plane PFM data for RBN is presented in **Figure 5** a) and b), respectively, where three main features are observed. Firstly, bright and dark regions, which correlate with the grains of the polycrystalline sample, are resolved. Secondly, grains with the strongest signal in the out-of-plane channel have typically a weaker signal in the in-plane channel and vice versa. For example, a grain with predominant in-plane PFM response is marked by a star in Figure 5a) and b), whereas another grain with a more pronounced out-of-plane PFM signal is marked by a cross. Lastly, two exceptions are outlined with a white-dotted line in Figure 5b). Here, the in-plane and out-of-plane contributions seem to be roughly equal. Still, with a line-trace across the grain boundary (see Figure 5c)), we notice a smooth transition in elevation between the two grains, while the relative angle of the piezoresponse, defined as $\arctan\left(\frac{PFM_{OOP}}{PFM_{IP}}\right)$, switches abruptly. The latter is important as it shows that the PFM contrast is indeed related to the local piezoelectric response and not governed by grain-to-grain variations caused by topography.

The PFM data is thus consistent with the polycrystalline nature of our samples and the piezoelectric response expected due to $Q_{piezo}$. For KBN, a piezoelectric response cannot be ruled out, but it is less clear due to a weaker signal-to-noise ratio, as shown in Figure S8. For the individual grains, a closer inspection of the PFM data in Figure 5 shows a homogeneous contrast distribution, with no sign of domain formation, as expected for piezoelectric systems that don't exhibit spontaneous polarization (*i.e.*, ferroelectric).[41] The local measurements thus further corroborate the non-ferroelectric nature of RBN and KBN, with piezoelectricity emerging as secondary effect of the non-centrosymmetric chiral structural order.

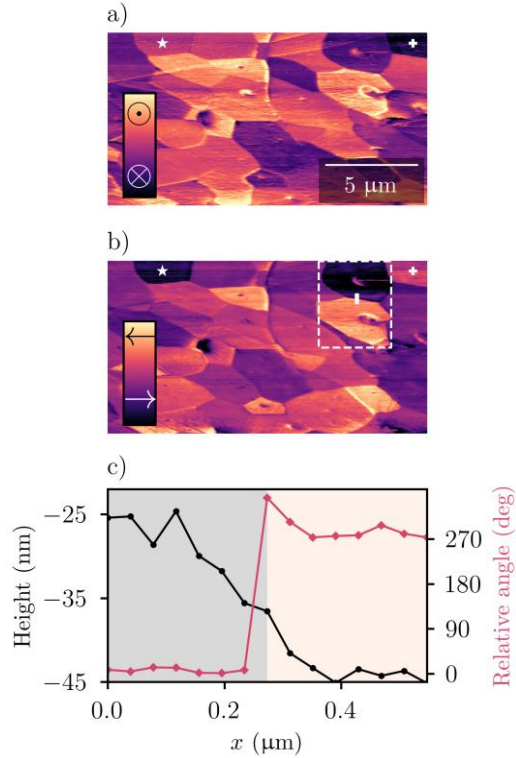

**Figure 5:** PFM of polycrystalline RBN showing clear a) out-of-plane and b) in-plane piezoresponse contrast between grains. c) The calculated relative angle (defined as **arctan** $(\frac{PFM_{OOP}}{PFM_{IP}})$) between the out-of-plane and in-plane response along the line inside of the dashed rectangle in b), showing two distinct signals.

## 3. Outlook

The results establish the TTB systems KBN and RBN as chiral materials, with an achiral-to-chiral phase transition that is driven by the chirality itself, representing the symmetry breaking primary order parameter. Importantly, and different from other systems reported in literature, the chiral order in KBN and RBN does not coexist with other types of ferroic order, such as ferroelectricity, making them prime candidates as pure ferrochiral materials. Such pure ferrochirality is desirable as it enables chirality control without crosstalk and unwanted contributions from other types of ferroic orders, which is of interest from an academic point of view, as well as chirality-based technological applications.

Intuitively, switching chirality is inherently challenging, as chemical bonds typically will need to be broken and reformed during switching. For example, an organic crystal consisting of only one enantiomer (one of a pair of chiral molecules), cannot be "chirally" switched without breaking and reforming bonds, as the "switched" enantiomer does not exist in the material. However, for KBN and RBN, the chirality is directly linked to the handedness of the helix so switching does not require breaking and reforming of any chemical bonds.

### 3.1. Proposed chiral switching field for KBN and RBN

Because of the piezoelectric property that follows with the chirality in KBN and RBN, we propose that a combined electric field and strain field, *i.e.*, an electric-strain field, can act as a chirality switching field. Invariant analysis shows that there exists an odd-order free energy term between the order parameters $Q_{piezo} \times Q_S \times Q_P$. Here, $Q_{piezo}$ is the piezoelectric order parameter transforming as $\mathbf{\Gamma_1^-}$ (*a*), $Q_S$ is symmetry breaking shear strain transforming as $\mathbf{\Gamma_5^+}$ (*b*,0) and $Q_P$ is the in-plane polarization order parameter transforming as $\mathbf{\Gamma_5^-}$ (*d*,0). This implies that an external electric-strain field can switch the sign of $Q_{piezo}$, which in turn links with, and changes the sign of $Q_A$ (the chiral order parameter transforming as $\mathbf{\Lambda_5}$ (*a*,0,0,0)), thus changing the handedness of the chiral structure, schematically shown in **Figure 6**. Both strain and electric fields interact more strongly with matter than, for example, light and could therefore ease a chiral switch to happen compared to previous attempts.[42]

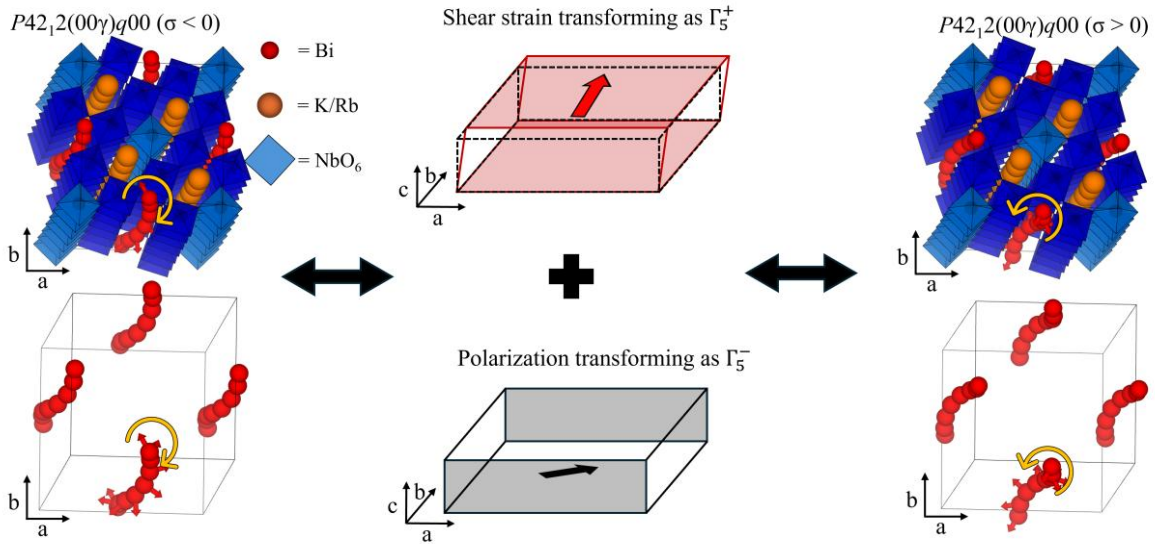

**Figure 6.** A 1x1x7 commensurate approximation of the incommensurate $P42_12(00\gamma)q00$ structure of KBN/RBN, showing the "right-" and "left-handed" version of the helix formed by $Bi^{3+}$ along the *c*-direction. Illustrated is also the shear strain transforming as $\mathbf{\Gamma_5^+}$ and the in-plane polarization transforming as $\mathbf{\Gamma_5^-}$ that in combination can be used to switch between the two chiral configurations.

Realization of the proposed chiral switching will have substantial implications for the application of (ferro)chiral materials and is an important next step in the research on chiral TTBs. It can be envisaged that ferrochiral TTBs can be used as switches and storage medium in photon based quantum computing and quantum networks.[7] This would then overcome two of the major limitations of state-of-the-art quantum computing, *i.e.*, operating at cryogenic temperatures and scalability.

## 4. Conclusion

The previously unsolved room temperature incommensurate structures of KBN and RBN have been solved by means of high-resolution X-ray powder diffraction. Both structures are described by superspace group $P4_22(00\gamma)q00$. We show that $Bi^{3+}$ takes on an unusual helical displacement, as we confirmed by atomic resolution STEM and explained by cooperative ordering of the stereochemically active $Bi^{3+}$ lone-pair. The chiral nature of such a helix, combined with the fact that space group $P4_22$ is one of the 43 non-enantiomorphic Sohncke space groups, makes the overall structure chiral.

Temperature dependent structural characterization reveals a phase transition from $P4/mbm$ to $P4_22(00\gamma)q00$ that can be classified both as commensurate-to-incommensurate and achiral-to-chiral phase transition. Symmetry analysis reveals that the primary order parameter of this transition is the chirality, transforming as $\boldsymbol{\Lambda_5}$ with order parameter direction ($a$, 0, 0, 0). A secondary order parameter, transforming as $\boldsymbol{\Gamma_1^-}$ with order parameter direction ($a$), is induced by this phase transition, making both KBN and RBN "improper piezoelectrics", as confirmed experimentally by PFM. Our structural characterization, combined with the PFM analysis, confirms the non-ferroelectric nature of the samples previously reported.

Finally, we propose both RBN and KBN as novel ferrochiral candidate materials. Both possess an achiral-to-chiral phase transition, and no other ferroic orders are known. Through invariant analysis we show that an electric-strain field (*i.e.*, combination of an electric field inducing in-plane polarization with a symmetry breaking shear strain) can be used to switch the handedness, and thus the chirality of the structure. Ferrochiral materials are highly sought after, as they would give direct external field control over all chirality related phenomena, *e.g.*, optical activity and chiral quantum optics.

## 5. Methods

KBN and RBN pellets were prepared via a two-step solid state synthesis route by first making $KNbO_3/RbNbO_3$ and $BiNbO_4$, then mixing $KNbO_3/RbNbO_3$, $BiNbO_4$ and $Nb_2O_5$ in stochiometric ratio and sintering at 1150 °C for 1h/8h. More details about the synthesis are given in ref.[24] The surfaces of the pellets were polished (~ 0.1 mm removed) before further processing.

*X-ray high-resolution powder diffraction:* Powders for X-ray powder diffraction were obtained by crushing and grinding the polished pellets, with subsequent sieving through a 25 μm mesh. The powders were filled in 0.5 mm diameter borosilicate capillaries. High-resolution X-ray powder diffraction data were collected in transmission geometry on the high-resolution setup at the ID22 beamline, European Synchrotron Radiation Facility (ESRF), Grenoble, France.[43, 44]. Data was collected at 4 K, 25 K, 50 K, 75 K and 100 K in a liquid helium cooled cryostat (5 data collections). A liquid nitrogen cryostream was used to control the temperature from 125 K to 475 K in 25 K steps (16 data collections). A hot air blower was used to control the temperature at 498 K and then every 50 K up to 798 K (723 K for KBN), with an additional data collection at 823 K (6 and 8 data collections for KBN and RBN, respectively). A silicon standard (NIST, 640c) was used to calibrate the energy (0.354265(3) Å) and instrumental contribution to peak broadening. The data was collected up to a Q around 11 Å$^{-1}$.

*Scanning transmission electron microscopy:* Samples for scanning transmission electron microscopy (STEM) were prepared by mixing KBN or RBN powders with fine aluminum powder in an approximate volume ratio of 1:10. This powder mixture was then sandwiched in aluminum foil, before being cold rolled down to approximate 100 μm thickness. The sample was then mechanically thinned to around 20 μm, before electron transparency was achieved through ion milling using Ar$^+$ ions at 6 keV. A final ion milling was performed with a beam energy of 100 eV to reduce surface damage. A JEOL 2100 LaB$_6$ transmission electron microscope (TEM) was used for initial imaging and selected area electron diffraction of the prepared samples, while an aberration corrected JEOL ARM200F TEM/STEM was used for acquiring the annular dark field (ADF) STEM data (both microscopes operated at 200 kV). Post acquisition, the micrographs were treated using a band-pass filter to reduce noise.

*Piezoresponse force microscopy:* The polycrystalline sample surfaces were prepared for piezoresponse force microscopy (PFM) by lapping using a 9 μm Al$_2$O$_3$ water suspension (Logitech, Glasgow, Scotland), followed by polishing utilizing a silica slurry (SF1 Polishing fluid, Logitech), leading to root mean square roughness of ± 20 nm for KBN and ± 33 nm for RBN. PFM was subsequently performed using the NTEGRA II system from NT-MDT (Spectrum Instruments Ltd., Limerick, Ireland) with an electrically conductive single crystal doped diamond tip (DEP01, TipsNano, Tallinn, Estonia). The instrument was operated off-resonance at 10 V with a frequency of 40.14 kHz. A periodically poled out-of-plane LiNbO$_3$ sample (PFM03, Spectrum Instruments Ltd., Limerick, Ireland) was used to calibrate the signals (out-of-plane and in-plane piezoresponse), which were read-out using two lock-in amplifiers (SR830, Standford Research Systems, CA, USA).

*Incommensurate modulated structure solution and refinement:* Rietveld refinements considering only the average aristotype structure were done using *TOPAS* (Bruker AXS, version 6).[45] Lattice parameters, scale factor, zero error, and Lorentzian and Gaussian strain contributions to the peak shape were always refined. The background was described by straight line segments connecting ~ 60 manually selected background points, with one overall scaling factor for all points being refined. The instrumental contribution to peak broadening was modelled with a *Voigt* peak-shape, for which parameters were fixed to the values obtained on a silicon standard. All 10 symmetry-allowed structural parameters were refined in addition to five thermal parameters (one for each cation site in the *P*4/*mbm* space group, four in total, and one for all oxygen sites).

These Rietveld refinements revealed significant discrepancies in the reflection intensities for both samples, see Figure S9. Allowing occupational disorder on the A-sites, which is a common feature of TTBs,[46-48] did not provide a satisfactory fit improvement. Investigating the difference Fourier maps revealed clear mismatch in the calculated and experimental electron densities on the A1/$Bi^{3+}$-site. It seems clear that $Bi^{3+}$ is off-centered away from the 2*a* Wyckoff site with coordinate (0, 0, 0). It is also clear that this off-centering is in-plane, suggesting that an improved refinement would be achieved by placing $Bi^{3+}$ on the 8*i* Wyckoff site with coordinate (*x*, *y*, 0). Allowing positional disorder of $Bi^{3+}$ on the A1-site significantly improved the fit, as shown in Figure S9 and reduced $R_{wp}$ from 20.29 to 14.14 and 24.00 to 15.86 for KBN and RBN, respectively. This is taken as a signature of the main structural distortion associated with the incommensurate structure, *i.e.*, the superspace group must allow for in-plane displacement of the A1-site.

*ISODISTORT* [49] was then used to generate potential superspace groups consistent with the average space group *P*4/*mbm* and the incommensurate modulation vector of the form [0, 0, γ] (*i.e.*, search over specific *k* points, namely Λ(0, 0, γ), in the first Brillouin zone). All but one candidate could be excluded by indexing and/or symmetry constraints of the modulations on the A1-site, *i.e.*, not allowing in-plane displacement (*i.e.*, *P*4/*mbm*(00γ)0000, *P*4/*mbm*(00γ)*s*0*s*0, *P*4/*mbm*(00γ)*s*00*s*, *P*4/*mbm*(00γ)00*ss* and *P*42$_1$2(00γ)000). This then left superspace group *P*42$_1$2(00γ)*q*00 (no. 90.1.19.2) with cell basis {(1,0,0), (0,1,0), (0,0,1)} and origin shift (1/2, 0, 0) compared to the *P*4/*mbm* parent structure as the only structural description. This was further confirmed by superspace Rietveld refinements performed using *Jana2020*.[50] *Jana2020* was also used to perform temperature dependent Rietveld refinements of the incommensurate modulated structure. These refinements were performed as explained above, with the addition of refining the modulation vector and all 36 structural parameters associated with the 1$^{st}$ order modulated displacements.

*Excluding possible polar distortions:* Although a polar distortion cannot be strictly ruled out based on the diffraction data in this work, one can show with *ISODISTORT* that neither of the polar subgroups of *P*4/*mbm* associated with order parameters transforming as $\Lambda_x$ ($x$ = 1, 2, 3, 4, 5) with the high symmetry order parameter direction (*i.e.*, of the form ($a$, 0)) are tetragonal (only orthorhombic or lower symmetries). These can readily be discarded based on the clear tetragonal symmetry observed in the high-resolution X-ray diffraction data. Allowing for a lower symmetry order parameter direction (*i.e.*, of the form ($a$, $b$)) *ISODISTORT* gives five polar tetragonal structures (*P*4*bm*(00γ)000, *P*4*bm*(00γ)*s*0*s*, *P*4*bm*(00γ)*ss*0, *P*4*bm*(00γ)0*ss* and *P*4(00γ)*q*), all of which can be excluded by indexing and/or symmetry constraints on the modulations on the A1-site, as described above. This then leaves the solved non-polar tetragonal structure, thus confirming the non-ferroelectric nature of KBN and RBN. This is consistent with previous work observing no ferroelectric hysteresis.[24]

**Acknowledgements**

O.G.G acknowledges financial support from NTNU (Norwegian University of Science and Technology). C.R.Z. and T.G. acknowledge funding from the Norwegian Research Council through the project High-Temperature Lead-Free Ferroelectrics based on Tungsten Bronzes (301954). R.S.D. and D.M. acknowledge funding from the European Research Council (ERC) under the European Union's Horizon 2020 Research and Innovation Program (Grant Agreement No. 863691). R.S acknowledges funding from EPSRC grant EP/V053701/1. M.S.S acknowledges the Royal Society for a fellowship (URF\R\231012). The European Synchrotron Radiation Facility (ESRF) is acknowledged for provision of beam time at ID22 (proposal number: hc5598, DOI: 10.15151/ESRF-ES-1578705079). The authors wish to acknowledge Dmitry Chernyshov, Giorgia Confalonieri, Donald Evans and Jere Tidey for fruitful discussion regarding data interpretation and analysis. Yining Xie is acknowledge for help in making the displacement vector/field maps.

**Author contributions**

O.G.G. initiated and planned the work, collected the X-ray diffraction data and performed the formal analysis in collaboration with M.S.S. C.R.Z. synthesized the samples and assisted in collecting the X-ray diffraction data under supervision by T.G. R.S.D. prepared samples for, and collected the piezoresponse force microscopy data under supervision by D.M. R.S. and R.B. collected and analyzed the scanning transmission electron microscopy data. O.G.G.

wrote the manuscript with critical feedback from all authors. All authors approved the final manuscript.

**Data Availability Statement**

X-ray diffraction data will be made available through the ESRF data portal after the embargo period (DOI: 10.15151/ESRF-ES-1578705079). Scanning transmission electron microscopy and piezoresponse force microscopy data that support the findings of this study are available within the article and its supplementary material.

# Supporting Information

**Chiral lone-pair helices with handedness coupling to electric-strain fields**

*Caren R. Zeiger[#], Ruben S. Dragland, Robin Sjökvist, Richard Beanland, Dennis Meier, Tor Grande, Mark S. Senn and Ola G. Grendal\**

C. R. Zieger, R. S. Dragland, D. Meier, T. Grande, O. G. Grendal
Department of Materials Science and Engineering, NTNU Norwegian University of Science and Technology, Sem Sælands vei 12, 7034 Trondheim, Norway
E-mail: ola.g.grendal@ntnu.no

R. Sjökvist, R. Beanland
Department of Physics, University of Warwick, Gibbert Hill Road, Coventry CV4 7AL, United Kingdom

M. S. Senn
Department of Chemistry, University of Warwick, Gibbert Hill Road, Coventry CV4 7AL, United Kingdom

[#] Current affiliation: SINTEF Industry, Richard Birkelands vei 3, 7034 Trondheim, Norway

**Rietveld refinement of RBN using superspace group $P42_12(00\gamma)q00$**

Result of the Rietveld refinement performed with *Jana2020* as described in the main is shown in **Figure S1** for RBN.

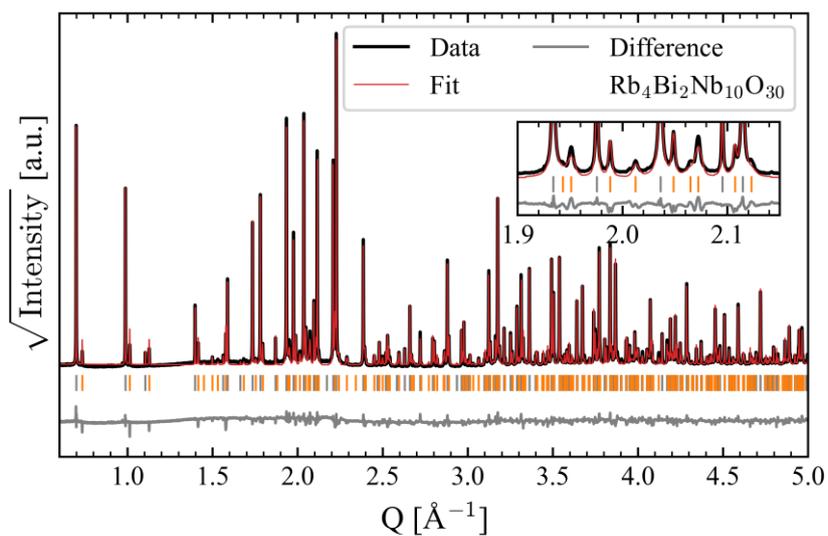

**Figure S1:** Rietveld refinement of RBN using superspace group $P42_12(00\gamma)q00$. Unit cell parameters: $a$ = 12.72100(2), $c$ = 3.95673(1), and $k$ = [0, 0, 0.14053(5)]. R-values: $R_{exp}$ = 2.33, $R_{wp}$ = 10.59, and $GOF$ = 4.55.

# Crystallographic data for the room temperature incommensurate structure of KBN and RBN

Crystallographic data from Rietveld refinements performed with *Jana2020* as described in the main text are given in **Table S1** and **Table S2** for KBN and RBN, respectively.

**Table S1:** Refined incommensurate room temperature structure of KBN using superspace group $P42_12(00\gamma)q00$. Atomic coordinates, amplitudes of the displacive modulation functions and $B_{iso}$ values. Values without uncertainties are fixed due to symmetry or instabilities in the refinements, while values starting with an "=" -sign are constrained by symmetry. Unit cell parameters: $a = 12.63982(3)$, $c = 3.92458(1)$ and $k = [0, 0, 0.1379(2)]$. R-values: $R_{exp} = 2.14$, $R_{wp} = 17.43$, $GOF = 8.14$. The displacive modulation terms are sorted by terms ("s" for sin and "c" for cos).

| Atom/Site/Wyckoff | | x [-] | y [-] | z [-] | $B_{iso}$ [Å$^2$] |
|---|---|---|---|---|---|
| Bi$^{3+}$/A1/2c | | 0 | 0.5 | 0.0189(7) | 1.92(4) |
| | s | 0.0186(2) | 0.0307(2) | 0 | |
| | c | =0.0307(2) | =-0.0186(2) | 0 | |
| K$^+$/A2/4e | | 0.1724(3) | =0.1724(3) | 0 | 1.1(1) |
| | s | -0.0040(7) | 0.0095(6) | -0.021(3) | |
| | c | =0.0095(6) | =-0.0040(7) | =0.021(3) | |
| Nb$^{5+}$/B1/2b | | 0 | 0 | 0.5 | 0.67(2) |
| | s | -0.0030(4) | 0.0064(3) | 0 | |
| | c | =0.0064(3) | =-0.0030(4) | 0 | |
| Nb$^{5+}$/B2/8g | | 0.0738(1) | 0.7162(1) | 0.5065(9) | =0.67(2) |
| | s | 0.0043(3) | 0.0115(2) | 0.007(1) | |
| | c | 0.0064(3) | 0.0014(3) | -0.0327(8) | |
| O$^{2-}$/O1/4f | | 0.2856(6) | =0.2856(6) | 0.5 | 0.1 |
| | s | -0.001(2) | -0.007(2) | -0.032(5) | |
| | c | =-0.007(2) | =-0.001(2) | =0.032(5) | |
| O$^{2-}$/O2/8g | | 0.1398(7) | 0.5654(7) | 0.486(5) | =0.1 |
| | s | -0.022(1) | -0.006(2) | 0.003(8) | |
| | c | -0.011(2) | 0.005(2) | -0.055(4) | |
| O$^{2-}$/O3/8g | | 0.9985(6) | 0.8431(6) | 0.532(5) | =0.1 |
| | s | 0.002(2) | -0.004(1) | 0.101(5) | |
| | c | 0.003(1) | -0.004(2) | -0.031(8) | |
| O$^{2-}$/O4/2a | | 0 | 0 | 0 | =0.1 |
| | s | -0.022(1) | -0.000(2) | 0 | |
| | c | =-0.000(2) | =-0.022(1) | 0 | |
| O$^{2-}$/O5/8g | | 0.0733(6) | 0.6968(6) | -0.027(4) | =0.1 |
| | s | -0.001(2) | -0.016(1) | 0.023(8) | |
| | c | 0.009(2) | 0.011(2) | -0.037(7) | |

**Table S2:** Refined incommensurate room temperature structure of RBN using superspace group $P4_22_12(00\gamma)q00$. Atomic coordinates, amplitudes of the displacive modulation functions and $B_{iso}$ values. Values without uncertainties are fixed due to symmetry or instabilities in the refinements, while values starting with an "=" -sign are constrained by symmetry. Unit cell parameters: $a = 12.72100(2)$, $c = 3.95673(1)$ and $k = [0, 0, 0.14053(5)]$. R-values: $R_{exp} = 2.33$, $R_{wp} = 10.59$, $GOF = 4.55$. The displacive modulation terms are sorted by terms ("s" for sin and "c" for cos).

| Atom/Site/Wyckoff | | $x$ [-] | $y$ [-] | $z$ [-] | $B_{iso}$ [Å²] |
|---|---|---|---|---|---|
| $Bi^{3+}$/A1/2c | | 0 | 0.5 | 0.0147(4) | 1.52(2) |
| | s | 0.0190(1) | 0.0348(1) | 0 | |
| | c | =0.0348(1) | =-0.0190(1) | 0 | |
| $Rb^+$/A2/4e | | 0.1735(1) | =0.1735(1) | 0 | 0.71(2) |
| | s | -0.0027(2) | 0.0047(1) | -0.0079(7) | |
| | c | =0.0047(1) | =-0.0027(2) | =0.0079(7) | |
| $Nb^{5+}$/B1/2b | | 0 | 0 | 0.5 | 0.50(1) |
| | s | -0.0057(2) | 0.0058(1) | 0 | |
| | c | =0.0058(1) | =-0.0057(2) | 0 | |
| $Nb^{5+}$/B2/8g | | 0.07257(5) | 0.71634(5) | 0.50934(5) | 0.50(1) |
| | s | 0.0045(1) | 0.0102(1) | 0.0139(6) | |
| | c | 0.0054(1) | 0.0023(2) | -0.0145(6) | |
| $O^{2-}$/O1/4f | | 0.2900(3) | =0.2900(3) | 0.5 | 0.1 |
| | s | -0.0014(9) | -0.0038(8) | -0.008(3) | |
| | c | =-0.0038(8) | =-0.0014(9) | =0.008(3) | |
| $O^{2-}$/O2/8g | | 0.1391(4) | 0.5681(4) | 0.505(3) | 0.1 |
| | s | -0.0163(8) | -0.0037(9) | 0.009(4) | |
| | c | -0.0019(9) | 0.0107(9) | -0.015(4) | |
| $O^{2-}$/O3/8g | | 0.9994(3) | 0.8433(4) | 0.504(4) | 0.1 |
| | s | -0.002(1) | -0.0054(7) | 0.059(2) | |
| | c | 0.0045(7) | 0.003(1) | 0.010(4) | |
| $O^{2-}$/O4/2a | | 0 | 0 | 0 | 0.1 |
| | s | -0.005(1) | -0.009(9) | 0 | |
| | c | =-0.009(9) | =-0.005(1) | 0 | |
| $O^{2-}$/O5/8g | | 0.0730(3) | 0.6970(3) | -0.002(4) | 0.1 |
| | s | 0.002(1) | -0.0069(8) | 0.0034(9) | |
| | c | -0.000(8) | -0.007(9) | -0.037(3) | |

**Selected area electron diffraction and atomic resolution scanning transmission electron microscopy (STEM)**

Atomic resolution STEM of RBN along the [120] zone axis is shown in **Figure S2**. Experimental details are found in the main text.

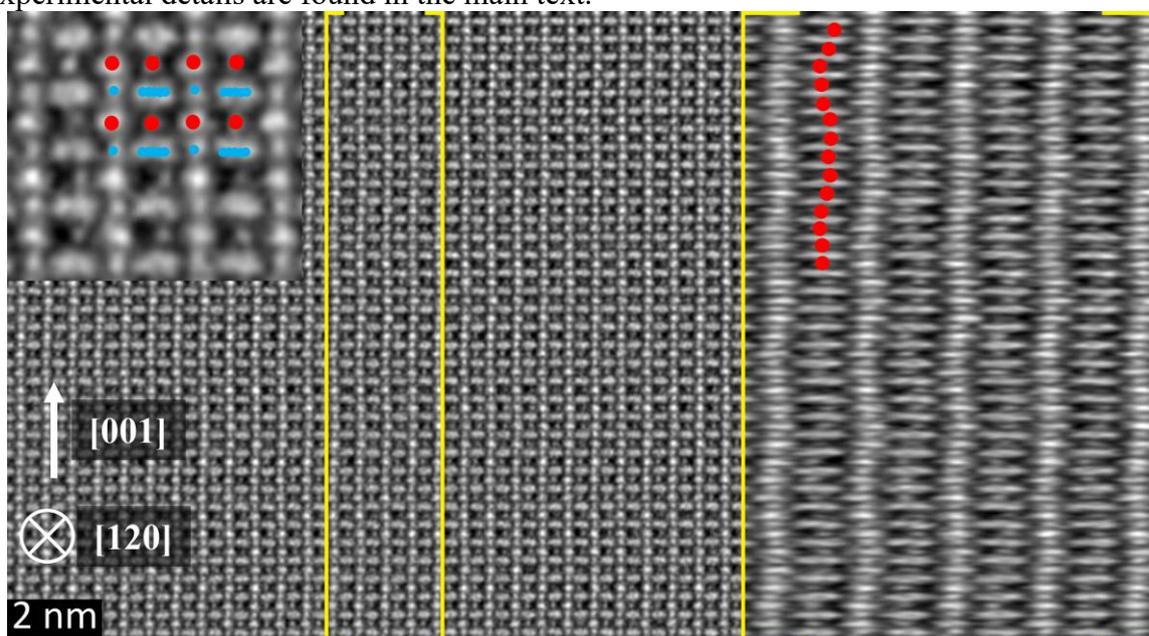

**Figure S2:** Atomic resolution STEM of RBN along the [120] zone axis. Insert shows the $Bi^{3+}$ (red) and $Nb^{5+}$ (blue) overlayed, and right panel is stretched to highlight the modulated displacement of $Bi^{3+}$ atoms

By analyzing the displacement of $Bi^{3+}$ atoms in relation to the surrounding atoms within the lattice, displacement vector maps can be constructed from atomic resolution STEM micrographs shown in **Figure S3** (a) and (d). These indicate both the magnitude and direction of the atomic displacements from a specific position within the lattice, here chosen as the center of the individual rectangles formed by the atoms surrounding the $Bi^{3+}$ sites (b) and (e). For increased visibility the vectors can then be translated into a radial color scheme, creating a displacement field map (c) and (f). This clearly illustrates how the $Bi^{3+}$ atoms form rows displaced either to the right or left in the figure, highlighting the helical displacement perpendicular to the *c* axis.

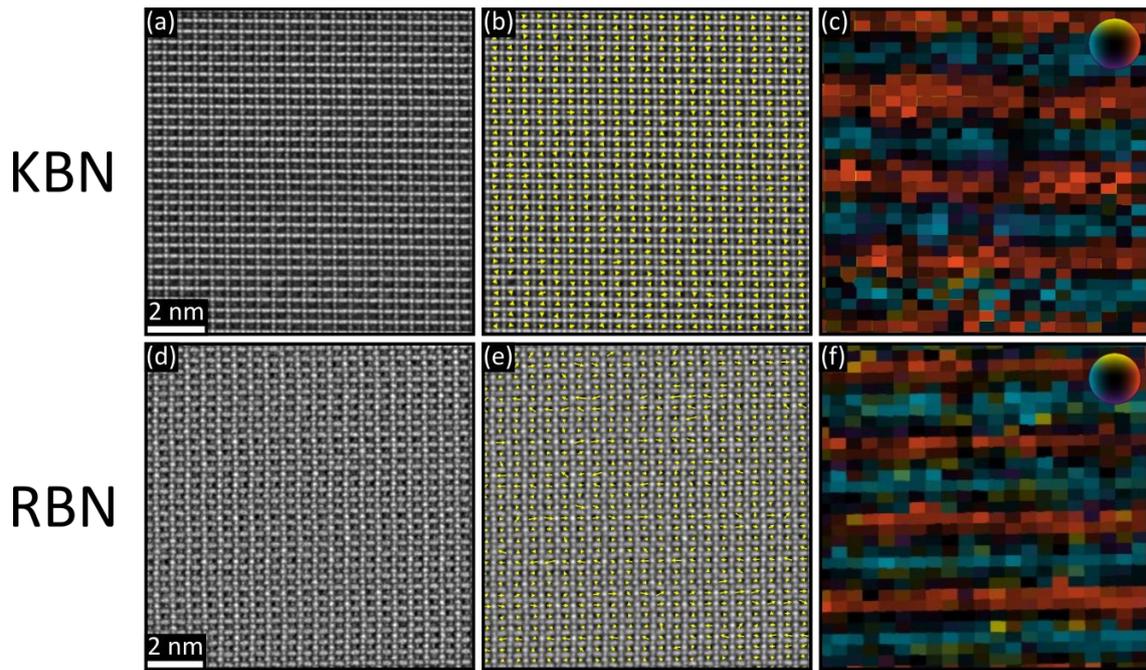

**Figure S3a:** Atomic resolution STEM of KBN (a) and RBN (d) along the [120] zone axis, with corresponding $Bi^{3+}$ displacements represented as vectors (b and e) and as fields (c and f).

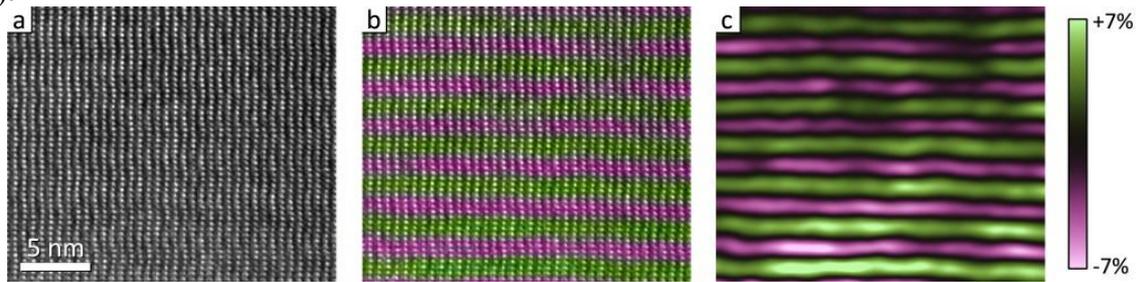

**Figure S3b:** a) High resolution TEM of KBN at the [110] zone axis, with scattering dominated by the $Bi^{3+}$ atoms showing a clear helical displacement (c-direction vertical). c) shows the $e_{xy}$ shear component of the image obtained using geometric phase analysis (GPA). b) is an overlaid composite of (a) and (c). Lower magnification TEM images showed the modulation to be consistent over lengths of at least 100s of nm.

Selected area electron diffraction patterns recorded along the [001] and [110] zone axes are presented in **Figure S4** for KBN and RBN. Clear satellite reflections that can be indexed with a modulation vector of the form [0, 0, γ] are observed for both KBN and RBN along the [001] direction in the [110] zone axis patterns. The corresponding incommensurate modulation vectors are found to be [0, 0, 0.133(1)] and [0, 0, 0.141(1)] for KBN and RBN, respectively.

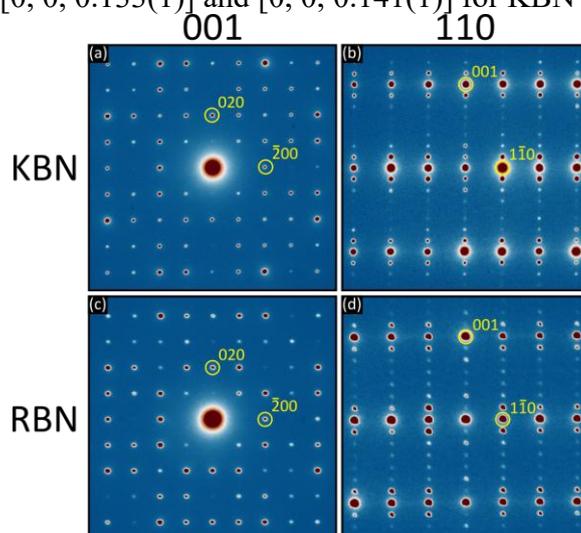

**Figure S4:** Selected area electron diffraction patterns for KBN (a and b) and RBN (c and d) along the [001] (a and c) and the [110] (b and d) zone axis. Clear satellite reflections are observed for both KBN and RBN.

In addition to the above-mentioned satellite reflections, an additional set of incommensurate satellite reflections were observed for KBN, shown in **Figure S5**. Dark field imaging reveals that this second set of satellite reflections are associated with a complex domain structure at a length scale of tens of nanometers, see Figure S5.

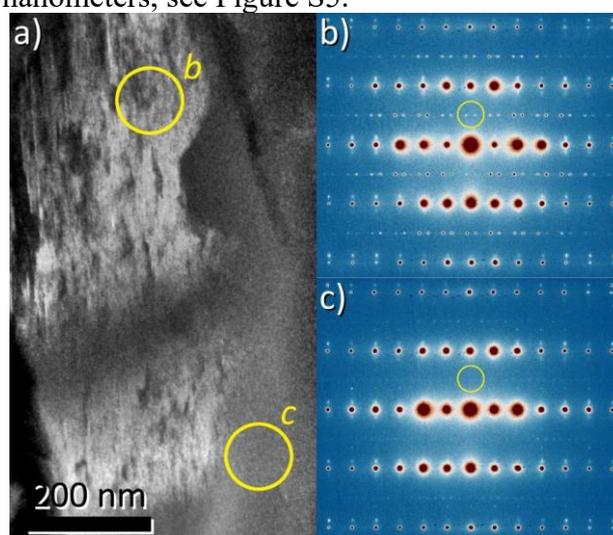

**Figure S5:** a) A dark field image of KBN obtained at the [772] zone axis ([-110] horizontal), using satellite spots selected by an objective aperture placed at the position marked by the yellow circles in b) and c). Bright regions in a) show the areas of the material where satellite reflections are present, and dark regions show areas where they are absent. The selected area electron diffraction patterns in b) and c) were obtained using a selected area aperture located in the regions indicated by yellow circles in a), showing the presence and absence of satellite reflections.

**Rietveld refinement of the high-temperature data of KBN and RBN using *P*4/*mbm***

The high-temperature X-ray diffraction data of both KBN and RBN shows no evidence for satellite reflections and is perfectly described by the aristotype space group *P*4/*mbm*. However, the data shows clear indications of positional disorder on the $Bi^{3+}$/A1-site. Rietveld refinements allowing for positional disorder are presented in **Figure S6** and show a great match with the experimental data. Other details about the Rietveld refinements are found in the main text. Detailed crystallographic data are given in Table S3 and S4 for KBN and RBN, respectively.

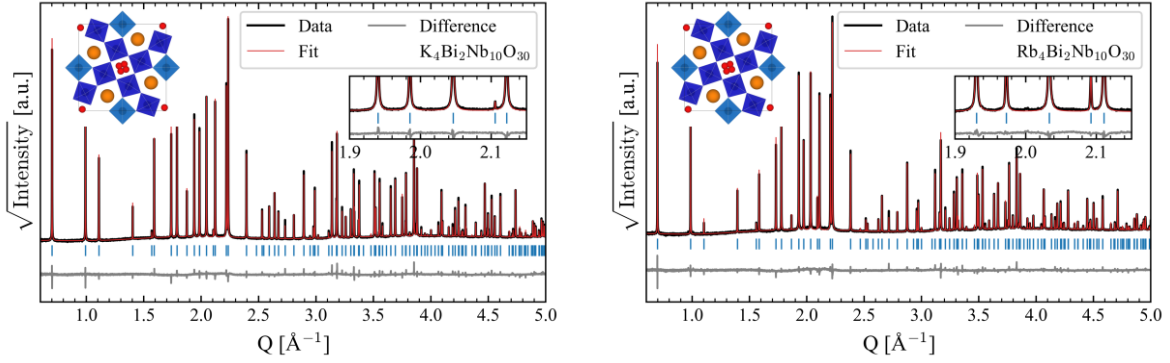

**Figure S6:** Rietveld refinements using *P*4/*mbm* and allowing for positional disorder on the $Bi^{3+}$/A1-site for KBN at 723 K (left) and RBN at 823 K (right).

**Table S3:** Refined structure of KBN using space group *P*4/*mbm* at 723 K. Atomic coordinates and $B_{iso}$ values are listed for the different atomic sites. Values without uncertainties are fixed due to symmetry, while values starting with an "=" -sign are constrained by symmetry or due to instabilities. Unit cell parameters: $a$ = 12.659979(6), $c$ = 3.948677(3). R-values: $R_{exp}$ = 2.15, $R_{wp}$ = 6.79, $GOF$ = 3.15.

| Atom/Site/Wyckoff | $x$ [-] | $y$ [-] | $z$ [-] | $B_{iso}$ [Å²] |
|---|---|---|---|---|
| $Bi^{3+}$/A1/8*i* | 0.0245(4) | 0.0129(7) | 0 | 5.00(3) |
| $K^+$/A2/4*g* | 0.1721(1) | =0.6721(1) | 0 | 3.44(4) |
| $Nb^{5+}$/B1/2*c* | 0 | 0.5 | 0.5 | 1.48(1) |
| $Nb^{5+}$/B2/8*j* | 0.07429(3) | 0.21611(3) | 0.5 | =1.48(1) |
| $O^{2-}$/O1/4*h* | 0.2873(2) | =0.7873(2) | 0.5 | 1.91(3) |
| $O^{2-}$/O2/8*j* | 0.3437(2) | 0.0022(2) | 0.5 | =1.91(3) |
| $O^{2-}$/O3/8*j* | 0.6365(2) | 0.4335(2) | 0.5 | =1.91(3) |
| $O^{2-}$/O4/2*d* | 0 | 0.5 | 0 | =1.91(3) |
| $O^{2-}$/O5/8*i* | 0.0747(3) | 0.2006(2) | 0 | =1.81(3) |

**Table S4:** Refined structure of RBN using space group $P4/mbm$ at 823 K. Atomic coordinates and $B_{iso}$ values are listed for the different atomic sites. Values without uncertainties are fixed due to symmetry, while values starting with an "=" -sign are constrained by symmetry or due to instabilities. Unit cell parameters: $a$ = 12.74031(1), $c$ = 3.965741(4). R-values: $R_{exp}$ = 2.30, $R_{wp}$ = 6.37, $GOF$ = 2.77.

| Atom/Site/Wyckoff | $x$ [-] | $y$ [-] | $z$ [-] | $B_{iso}$ [Å$^2$] |
|---|---|---|---|---|
| Bi$^{3+}$/A1/8$i$ | 0.0277(3) | 0.0139(5) | 0 | 5.00(3) |
| Rb$^+$/A2/4$g$ | 0.1738(1) | =0.6738(1) | 0 | 2.67(2) |
| Nb$^{5+}$/B1/2$c$ | 0 | 0.5 | 0.5 | 1.62(1) |
| Nb$^{5+}$/B2/8$j$ | 0.07278(3) | 0.21693(2) | 0.5 | =1.62(1) |
| O$^{2-}$/O1/4$h$ | 0.2909(2) | =0.7909(2) | 0.5 | 1.81(3) |
| O$^{2-}$/O2/8$j$ | 0.3440(2) | 0.0016(2) | 0.5 | =1.81(3) |
| O$^{2-}$/O3/8$j$ | 0.6359(2) | 0.4334(2) | 0.5 | =1.81(3) |
| O$^{2-}$/O4/2$d$ | 0 | 0.5 | 0 | =1.81(3) |
| O$^{2-}$/O5/8$i$ | 0.0731(2) | 0.2000(2) | 0 | =1.81(3) |

**Temperature dependent structural changes**

Rietveld refinement with a symmetry mode approach was used to extract temperature dependent structural changes. As mentioned, the phase transition from *P4/mbm* to *P4$_1$2(00γ)q*00 is driven by the primary order parameter $\Lambda_5$ with a single secondary order parameter transforming as the irreducible representation $\Gamma_1^-$. The temperature evolution of the displacive distortion amplitudes associated with $\Lambda_5$ and $\Gamma_1^-$ is presented in **Figure S7**, and especially $\Lambda_5$ shows a rapid increase below the phase transition before plateauing, indicative of a second-order/continuous phase transition.

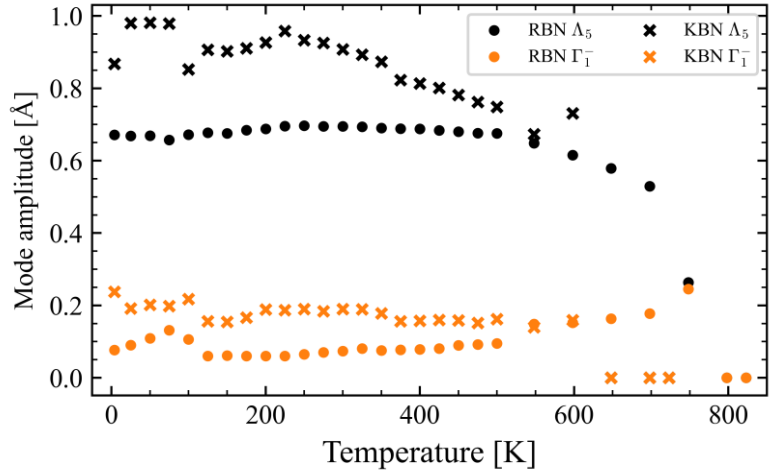

**Figure S7:** Temperature dependence of the displace distortion amplitudes associated with $\Lambda_5$ and $\Gamma_1^-$ obtained from Rietveld refinement for KBN and RBN.

**Piezoresponse force microscopy (PFM) of polycrystalline KBN**

PFM measured on KBN shows the same results as presented for RBN in the main text, although with a weaker signal-to-noise ratio, making the contrast less clear. Still, contrast between grains is observed, with inversion in contrast between in-plane and out-of-plane contributions. Also, no evidence for ferroelectric domains is observed.

In PFM, one would expect intergranular contrast with varying degree of in-plane and out-of-plane measured piezoresponse for a piezoelectric polycrystal. A dominant out-of-plane signal can be attributed to a dominant out-of-plane piezoresponse, but also to buckling caused by the in-plane response being parallel to the cantilever. Because of random crystallographic/polar orientations in polycrystals, some grains can exhibit equal piezoresponse in the out-of-plane and in-plane channel

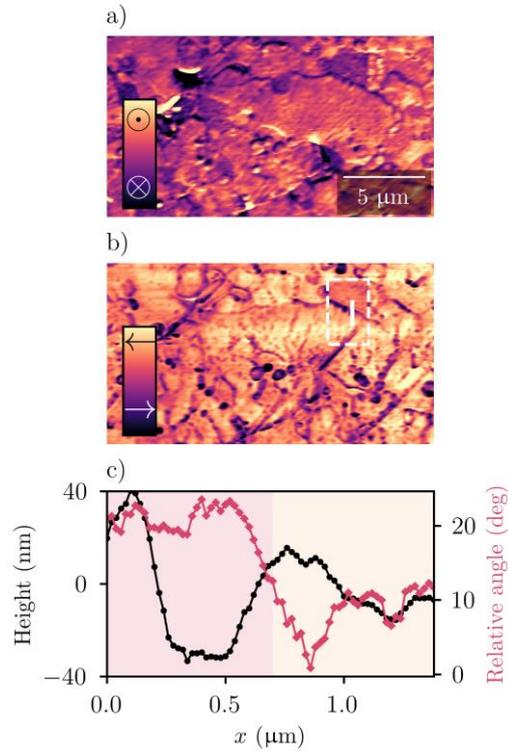

**Figure S8:** PFM of polycrystalline KBN showing a) out-of-plane and b) in-plane piezoresponse contrast between grains. c) The calculated relative angle (defined as $rel.\,angle = \arctan\left(\frac{PFM_{OOP}}{PFM_{IP}}\right)$) between the out-of-plane and in-plane response along the line inside of the dashed rectangle showing two distinct signals.

**Rietveld refinement using the average *P4/mbm* structure**

Rietveld refinement considering only the main reflections (*i.e.* disregarding the clear satellite reflections) revealed significant misfit in the intensities of the reflections for both KBN and RBN. Refinements were done as described in the main text. In **Figure S9** the Rietveld refinements compare a positionally ordered (*i.e.* $Bi^{3+}$ positioned at the *2a* Wyckoff site with coordinate (0, 0, 0)) and disordered (*i.e.* $Bi^{3+}$ position at the *8i* Wyckoff site with coordinate (*x*, *y*, 0)) structural model in *P4/mbm* is shown. Refined values for *x* and *y* for KBN and RBN are presented in Table S5, and *R*-values are given in Table S6. The positionally disordered model results in significant improvements in the Rietveld refinements.

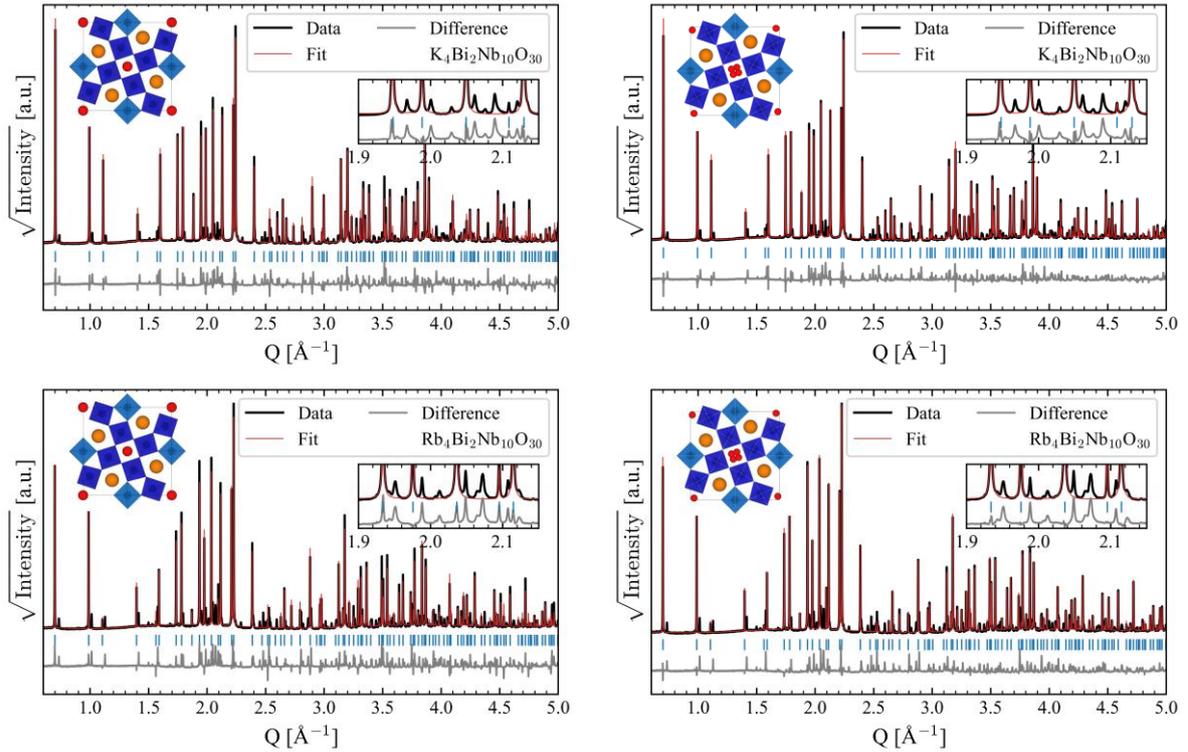

**Figure S9:** Rietveld refinements of KBN (top row) and RBN (bottom row) using the average *P4/mbm* space group with (left) and without (right) positional disorder.

**Table S5:** Refined fractional values for *x* and *y* for $Bi^{3+}$/*8i* Wyckoff site for positionally disordered RBN and KBN in the *P4/mbm* space group.

|     | *x* [-]   | *y* [-]   |
| --- | --------- | --------- |
| RBN | 0.0359(2) | 0.0192(3) |
| KBN | 0.0329(2) | 0.0155(4) |

**Table S6:** R-values for the room temperature Rietveld refinements with and without positional disorder for RBN and KBN. *Goodness of fit* (GOF) is defined as $R_{wp}/R_{exp}$.

|               | $R_{exp}$ [-] | $R_{wp}$ [-] | GOF [-] |
| ------------- | ------------- | ------------ | ------- |
| RBN, order    | 1.04          | 24.00        | 23.11   |
| RBN, disorder | 1.04          | 15.86        | 15.28   |
| KBN, order    | 1.02          | 20.29        | 19.97   |
| KBN, disorder | 1.02          | 14.14        | 14.19   |